# Ethical Risk Assessment of the Data Harnessing Process of LLM supported on Consensus of Well-known Multi-Ethical Frameworks

Javed I. Khan [†, *], Sharmila Rahman Prithula [†]

[†] Department of Computer Science, Kent State University, Ohio, USA

**Abstract**

The rapid advancements in large language models (LLMs) have revolutionized natural language processing, unlocking unprecedented capabilities in communication, automation, and knowledge generation. However, the ethical implications of LLM development, particularly in data harnessing, remain a critical challenge. Despite widespread discussion about the ethical compliance of LLMs—especially concerning their data harnessing processes—there remains a notable absence of concrete frameworks to systematically guide or measure the ethical risks involved. In this paper we discuss a potential pathway for building an Ethical Risk Scoring (ERS) system to quantitatively assess the ethical integrity of the data harnessing process for AI systems. This system is based on a set of assessment questions grounded in core ethical principles, which are, in turn, supported by commanding ethical theories. By integrating measurable scoring mechanisms, this approach aims to foster responsible LLM development, balancing technological innovation with ethical accountability.

## 1. Introduction

Large language models (LLMs) have emerged as transformative tools in artificial intelligence (AI), enabling applications across diverse domains such as healthcare, education, business, and entertainment [1,3,4]. Their ability to process and generate human-like text at scale is the result of vast training datasets and sophisticated training algorithms [6]. However, the reliance on extensive and often sensitive data raises significant ethical concerns, particularly in the areas of data acquisition, processing, and usage [2]. Ethical lapses in these processes can result in harm to individuals, organizations, infrastructure, animals, and the environment, amplifying the need for systematic mitigation strategies [6,8].

Current approaches to addressing ethical issues in AI often lack measurable frameworks that account for the many dimensions of potential harm. This gap highlights the necessity of embedding ethics as a fundamental principle in LLM development, from the initial data sourcing to the final model deployment. LLMs can perpetuate and amplify societal biases present in their training data, generate misinformation [4], and even infringe on privacy [1,2,10]. LLM-generated content can have detrimental effects on animals and the environment, through applications that impact animal and environmental welfare [8]; on organizations, through the spread of misinformation or data breaches [10]; and on infrastructure, through potential vulnerabilities introduced by malicious use. To address these challenges, it is necessary to evaluate both the data used to train LLMs and the process by which the data is harnessed. While ethical compliance in LLM data harnessing processes is widely discussed [6], there are few established methods that quantitatively assess the ethical risks associated with these processes.

The ethical dimensions of LLMs are complex, touching upon various established ethical theories, including consequentialist ethics, deontological ethics, virtue ethics [11,12], as well as considerations of justice, fairness, and care [13]. These theories provide a critical lens

Cite this article as: Khan, J., & Prithula, S. R. (2025). *Ethical Risk Assessment of the Data Harnessing Process of LLM supported on Consensus of Well-known Multi-Ethical Frameworks* . Proceedings of the 38th Canadian Conference on Artificial Intelligence, 2025, Calgary, Canada. Retrieved from https://caiac.pubpub.org/pub/s50xrmxh

through which the responsible development of LLMs can be assessed. In this paper, we propose a comprehensive framework based on these ethical theories for identifying and quantifying ethical risks associated with LLM training data. We introduce an Ethical Risk Scoring (ERS) mechanism that quantifies ethical vulnerabilities across four key metrics: harm potential, ethical sourcing, transparency, and target rights compliance. By integrating measurable scoring mechanisms, the ERS aims to foster responsible LLM development, balancing technological innovation with ethical accountability, while minimizing harm and maximizing societal benefit.

## 2. Related Work

The rapid advancement of LLMs has sparked widespread interest, not only in their transformative capabilities but also for the ethical and societal challenges they present. Prior research has extensively explored the vulnerabilities, risks, and mitigation strategies associated with LLMs, focusing on areas such as its harm potential to privacy, transparency, and ethical acquisition. Despite these advancements, there remains a lack of standardized, quantitative methodologies for evaluating ethical risks during data collection and preprocessing, particularly those related to societal bias, emotional harm, rights of the subject of data or target. Without quantifiable measures, ethical quality remains an aspiration rather than a standard practice. Only when ethics can be evaluated systematically can it meaningfully guide technological progress. This research aims to address ethical concerns related to privacy, consent, and representational diversity in training data. Although there is no significant work on scoring ethical quality, NIST's CVSS, developed to assess common vulnerabilities in software systems [19], provides a technical roadmap.

### 2.1. Common Ethical Concerns in Data Harnessing

The data used to train LLMs is a critical point of scrutiny, as it directly impacts the ethical integrity of the resulting systems [3,13,14]. Studies such as Bender et al. [5] and Gebru et al. [9] highlight the importance of transparency in dataset creation, supporting practices like documenting data origins, purposes, and potential.

### 2.2. Harm Dimensions

The literature identifies several harmful dimensions associated with LLMs, including social, informational, mental, environmental, and even animal welfare impacts.

#### 2.2.1. Social and Informational Harm

LLMs are susceptible to amplifying societal biases embedded in their training data, leading to representational harm and discrimination [1]. Furthermore, their ability to generate misinformation poses significant risks to tasks like open-domain question answering [2]. Proposed mitigations include robust monitoring systems, prompt engineering, and multi-reader verification methods to reduce bias and misinformation.

#### 2.2.2. Mental Harm

LLMs can lead to user manipulation or dependency, which may undermine user autonomy [1,6]. Addressing this requires designing systems that clearly communicate their limitations, educating users about safe interactions, and incorporating safeguards to prevent over-reliance on AI-generated content.

#### 2.2.3. Environmental Harm

Training large-scale models consumes significant computational resources, contributing to carbon emissions and ecological damage [4,6,14,15]. The literature emphasizes the broader environmental consequences, including water and soil degradation, advocating for energy-efficient architecture and investments in renewable energy to offset these impacts. Environment can be impacted by LLMs if used unwisely.

#### 2.2.4. Animal Welfare

While often overlooked, LLMs can indirectly harm animals by enabling technologies that exacerbate wildlife monitoring misuse or factory farming practices [4,8]. Expanding ethical frameworks to include non-human stakeholders is crucial for addressing this dimension.

## 2.3.    Transparency in AI

Transparency is widely regarded as essential for fostering trust and accountability in AI systems. Frameworks like "datasheets for datasets" (Gebru et al. [9]) have been proposed to improve clarity about a system's design, intended use, and limitations. However, these tools primarily focus on the deployment phase, leaving significant gaps in transparency during earlier stages like data harnessing and model training.

Broader frameworks, such as those proposed by [4], emphasize principles of beneficence, non-maleficence, and justice in AI design. These principles guide the development of transparent systems that respect the rights and well-being of all stakeholders, including humans, animals, and the environment.

Despite the progress in identifying vulnerabilities and proposing mitigation strategies, the field lacks an integrated approach to assessing the ethical robustness of LLMs. Current discourses significantly focus on data privacy but often overlook many other harms—such as potential social, informational, environmental, and animal welfare impacts. This highlights the need for tools like the Ethical Risk Scoring (ERS) Index, which can systematically evaluate harm potential, transparency, and sourcing integrity using diverse ethical principles. The ethical considerations surrounding LLMs require multidimensional, proactive approaches that address risks throughout the model development lifecycle. By synthesizing insights from harm mitigation, transparency frameworks, and ethical principles, this research aims to bridge the gaps in current methodologies and contribute to the development of more ethically robust LLM.

## 3.  Problem Definition

Rapid advances in machine learning (ML), especially with LLMs, have transformed how data is processed, generated, and used [1,3]. However, the growth and efficacy of these models rely significantly on massive volumes of training data, which raises numerous ethical concerns [2,5]. These issues are frequently overlooked during the development process, which can have severe consequences for society, organizations, and the environment [5,14].

This section discusses the major ethical concerns in LLM training data and explores the need for strong frameworks and evaluation methodologies to handle these issues consistently.

### 3.1.    Ethical Issues in LLM Training Data

The training data for LLM systems often originates from diverse, unregulated, or insufficiently vetted sources, leading to ethical dilemmas spanning several dimensions [9,13,14]. These issues impact individual rights, organizational integrity, societal values, and even environmental and animal welfare.

#### 3.1.1.  Bias and Representational Harm

ML training data often reflects societal biases, such as gender, racial, or cultural prejudices, due to skewed sampling or implicit stereotypes in data collection [14,]. For instance, biased datasets can lead to unfair treatment or discrimination in automated decision-making systems, disproportionately affecting vulnerable groups [2,9,14].

#### 3.1.2.  Data Privacy Violations

Data privacy violations occur when personal or sensitive information is used in training datasets without adequate safeguards, exposing individuals to risks of misuse or exploitation [3, 10, 14]. Such practices often bypass principles of data ownership, leaving individuals unaware that their data has been collected, processed, or monetized for machine learning (ML) systems [7,17].

#### 3.1.3.  Unethical Data Sourcing

Unethical data sourcing involves acquiring training data from sources that violate copyright laws, exploit unpaid labor, or scrape information without user consent [7,9,13]. This raises significant ethical concerns, particularly when data subjects and content creators are neither acknowledged nor compensated, despite their contributions fueling profit-driven AI systems [7,10].

### 3.1.4. Transparency and Accountability

Many training datasets lack documentation that provides visibility into their origins, processing, and curation [5,9,13,14]. This lack of transparency limits researchers' ability to identify and mitigate risks, leaving ethical concerns unresolved and diminishing accountability in LLM development [3].

## 3.2. Harm Potential in LLM Training Data

Harm resulting from the use of LLM training data manifests in various dimensions, including harm to humans, animals, infrastructure, organizations, and the environment [2,5,8]. The potential for harm arises from inappropriate data collection, unethical algorithmic applications, or mismanagement of sensitive information [3,6]. Each dimension of harm has unique characteristics, consequences, and mitigation requirements, necessitating a comprehensive approach to ethical safeguards in LLM systems [2,6,10,14].

### 3.2.1. Harm to Humans

Harm to humans can occur across multiple domains, including physical, emotional, cognitive, and economic well-being. Physical harm includes death, disability, illness, reduced quality of life, and even forms of enslavement, such as an LLM generating misleading medical advice may result in individuals following harmful practices, such as taking incorrect medications or delaying necessary medical interventions [6,14]. Emotional and psychological harm includes damage to reputation, distress, anxiety, and cyberbullying, which can be amplified by generating outputs that are harmful, such as offensive language, bullying, or spreading falsehoods about individuals [3,5,15]. To mitigate these risks, LLM systems must incorporate robust safeguards, including ethical data governance, algorithmic transparency, and mechanisms to prevent the misuse of data that directly impacts human well-being.

### 3.2.2. Harm to Animals

Animals are often subjected to harm caused by human activities, including physical abuse, psychological distress, and environmental destruction. LLMs used in industries like factory farming could inadvertently encourage harmful practices through biased recommendations or inaccurate assessments of humane methods [8]. Additionally, harm extends to species-level issues, including habitat loss, disruption of breeding patterns, genetic manipulation, overharvesting, and invasive experimentation [8,15]. Mitigating this harm requires integrating ethical AI guidelines, protecting sensitive ecological data, enforcing animal welfare laws, and designing AI systems that prioritize animal conservation and welfare.

### 3.2.3. Harm to Infrastructure and Services

Harm to infrastructure includes physical damage, unavailability of critical services, and integrity breaches. Critical infrastructure systems relying on LLMs for maintenance or operations could face downtime due to errors in LLM outputs.

### 3.2.4. Harm to Organizations:

Organizations face risks that range from economic losses to reputational damage, operational inefficiencies, and legal consequences. Misuse of LLMs could lead to financial losses for organizations [5,10]. Bias or harmful outputs from an LLM deployed in customer service or marketing could tarnish a company's reputatio.

### 3.2.5. Harm to the Environment:

Environmental harm includes pollution, climate change, deforestation, biodiversity loss, and waste accumulation. These issues are often interrelated, with cascading effects on ecosystems and human health. Improper use of ML systems can contribute to environmental harm through inaccurate resource management, excessive energy consumption from poorly optimized models, or overexploitation due to flawed predictions [5,6,8,15]. Mitigation requires adopting sustainable development goals in ML applications, ensuring data reliability, promoting energy-efficient AI practices, and enhancing transparency in environmental decision-making processes. These ethical issues are compounded by the absence of standardized frameworks for evaluating and mitigating risks within ML training data. While ethical guidelines exist, their implementation remains inconsistent, leaving significant gaps in accountability and ethical robustness.

## 4. Ethics And AI

One of the key challenges in developing a scoring system for ethics is its perceived subjectivity. To address this concern, we propose grounding the evaluation process in well-established ethical theories. Fortunately, the field of ethics offers an extraordinarily rich array of theoretical frameworks, each contributing a distinct lens through which to examine contemporary issues in data ethics. This diversity enables a comprehensive, 360-degree perspective on the moral dimensions of AI systems. Ethical theories provide structured guidance for evaluating critical considerations such as harm mitigation, fairness, transparency, and the protection of individual rights. By leveraging multiple normative frameworks, we can conduct a multidimensional assessment that ensures alignment between AI development and broader societal values.

### 4.1. Ethical Theories

These rich sets of theories can serve as a foundational tool for embedding ethical principles into AI system design and policy, fostering a thoughtful balance between technological innovation and social responsibility [3,11]. Here is a quick review of a few:

1) <u>Utilitarianism:</u> Utilitarianism, introduced by Jeremy Bentham and John Stuart Mill, emphasizes maximizing overall happiness while minimizing suffering. This framework evaluates the morality of actions based on their consequences, prioritizing the greatest good for the greatest number [3,12]. In the context of AI, this perspective could guide decisions such as designing systems that benefit a large population, such as AI-enabled universal healthcare systems, where societal welfare is prioritized.

2) <u>Deontological Ethics</u>: Deontological ethics, rooted in the works of Immanuel Kant, focuses on adherence to universal moral duties and rules [12]. This framework asserts that ethical actions must align with moral laws that could be applied universally, regardless of their consequences [3,11]. For example, in AI development, maintaining transparency and honesty in data processing—even when obscuring details might yield better results—aligns with deontological principles. The "categorical imperative" serves as a guiding principle, ensuring that AI systems operate in accordance with universal moral laws [12].

3) <u>Virtue Ethics</u>: Virtue ethics, championed by Aristotle, centers on the development of moral character. It suggests that ethical behavior emerges from cultivating virtues such as honesty, courage, and integrity [11,12]. In the context of AI, this framework emphasizes designing systems that reflect virtuous behavior, such as an AI healthcare system demonstrating compassion and honesty in its interactions with patients.

4) <u>Ethics of Care</u>: The ethics of care, introduced by Carol Gilligan, shifts the focus from rules and consequences to interpersonal relationships and empathy. This framework argues that ethical decisions must prioritize care, connection, and relational contexts. For instance, an AI system designed to assist in elder care would align with this framework by prioritizing the emotional and physical well-being of its users over rigid cost-efficiency metrics.

5) <u>Rights-Based Ethics</u>: John Locke's rights-based ethics emphasizes the inherent rights of individuals, such as freedom, equality, and privacy. This framework is particularly relevant in AI, where respecting the data rights of individuals, free of bias are paramount [7]. For example, AI developers must ensure that user privacy is protected, and systems are designed to prevent bias and unauthorized access to sensitive information.

6) <u>Social Contract Theory</u>: Social contract theory, rooted in the works of Thomas Hobbes and John Rawls, suggests that ethical behavior arises from agreements between individuals to benefit society [11]. This framework views morality as a product of collective consensus, emphasizing the importance of cooperation and mutual respect.

7) <u>Rawlsian Justice</u>: John Rawls' concept of justice as fairness emphasizes impartiality in decision-making to ensure equitable outcomes [7]. Using the "veil of ignorance," Rawlsian justice encourages decision-makers to design systems without knowledge of their own societal position, ensuring fairness for all, particularly the most disadvantaged. For instance, AI algorithms can be designed to prioritize equitable resource distribution, such as improving healthcare access for underprivileged communities.

8) <u>Natural Law Theory</u>: Natural Law theory, articulated by Thomas Aquinas, asserts that morality is grounded in universal truths derived from human nature and reason. This framework argues that ethical principles should align with natural rights and human purpose.

9) <u>Environmental Ethics</u>: Environmental ethics, championed by Aldo Leopold, expands the scope of morality to include the environment. It emphasizes sustainability and the ethical responsibility of humans to preserve and protect nature.

10) <u>Pragmatism:</u> Pragmatism, as articulated by William James and John Dewey, advocates for a focus on practical outcomes, adaptability, and effectiveness. This framework suggests that ethical decisions should be guided by their real-world utility and problem-solving capacity. For example, pragmatism in AI might involve implementing renewable energy solutions that balance economic viability with environmental preservation, demonstrating both practicality and ethical consideration.

## 4.2.    Ethical Principles

Ethical principles play a crucial role in guiding the design and development of AI systems, ensuring that they operate responsibly and transparently. These principles address key areas such as harm mitigation, data ownership, and the rights of data-subjects [10,12]. Derived from ethical theories, they provide a foundation for creating AI systems that prioritize the well-being of individuals, organizations, and the environment. The key ethical principles guiding the ethics of LLM data harnessing are catalogued in table 1.

4.2.1. Principles of Harm Mitigation

AI systems must be designed to minimize harm in all its forms. Principles H1 to H5 emphasize avoiding harm to individuals, organizations, infrastructure, animals, and the environment [6,8,15,17]. Designers are also responsible for taking necessary measures to mitigate potential harm (H6) [6,8,10] and disclose any residual harm transparently (H7) [14,16]. Additionally, while intentional harm is considered more unethical than unintentional harm (H8), the latter is still unacceptable (H9) [8]. The severity of harm must be prioritized, with harm to individuals being the most critical, followed by harm to the environment, animals, and organizations (H10) [15].

4.2.2. Principles of Data Ownership

Respecting data ownership is fundamental to ethical AI. Principles K1 to K3 stress that data must be acquired from its rightful owner, distributed in accordance with granted rights, and used in ways that uphold the rights of data subjects [7,15,17]. This ensures that data handling processes are both transparent and ethical, fostering trust in AI systems.

4.2.3. Principles of Subject Rights

Data subjects are entitled to a range of rights, ensuring transparency, consent, and control over their data. Principles S1 to S9 emphasize the subject's right to know how their data is collected and used, correct errors, withdraw consent, and request data erasure [7,10,15]. Furthermore, subjects have the right to restrict automatic decision-making (S10) and prevent the use of their data for training AI systems (S11) [16]. Subjects also have the right to prevent their data from being sold to third parties (S12) and should not face discrimination for exercising these rights (S13). Lastly, subjects are entitled to compensation if their data is monetized (S14), ensuring accountability and fairness in AI systems.

| P-Code | Ethical Principles |
|---|---|
| H1 | The system should do no harm to any individual. |
| H2 | The system should do no harm to an organization. |
| H3 | The system should do no harm to and infrastructure of system. |
| H4 | The system should do no harm to animals. |
| H5 | The system should do no harm to environment. |
| H6 | The system designer should take necessary measures within its control to mitigate the harm. |
| H7 | The system should disclose the net harm potential. |
| H8 | Intentional harm is more unethical than unintentional harm. |
| H9 | Even the unintentional harm is unethical. |
| H10 | The severity of harm is ordered as per harm to individual, environment, animal and organization. |
| K1 | Data should be acquired from the proper owner of the data. It should not be stolen. |
| K2 | Data should be distributed based on the distribution rights granted by the owner (alteration, commercial/noncommercial, attribution, purpose) |
| K3 | The rights of the data subject should be upheld by the data owner. |
| S1 | Any subject has the right to know if data being collected about him/her, what data is being collected, by whom, and for what purpose. |
| S2 | Any subject has the right to know the process of how his/her data is being collected. |
| S3 | Data should not be used for purposes not authorized by the subject. |
| S4 | Any subject has the right to be informed promptly if their data is used for purposes beyond their initial consent, allowing them to restrict such use. |
| S5 | Any subject has the right to withdraw permission for the use of their data later. |
| S6 | Any subject has the right to request copies and download the data about him/her in a readable format. |
| S7 | Any subject has the right to request correction of wrong information in the data about him/her. |
| S8 | An individual has the right to be forgotten i.e. erasure of his/her data |
| S9 | Any subject has the right to restrict how his/her data is being collected, processed, archived, or distributed and optout. |
| S10 | Any subject has the right to not be subject to automatic decision making. |
| S11 | Any subject has the right not to be subject to be used for AI system training. |
| S12 | Any subject has the right to restrict his/her information not to be sold to third party. |
| S13 | Any subject has the right to not be discriminated against for exercising his data rights by denial and restriction of other services. Individuals cannot be denied goods or services, or offered different prices or quality, for exercising their data rights |
| S14 | Any subject has the right to be compensated if any data about him is monetized profit. |

*Table 1 Principles underlying Ethical frameworks*

By adhering to these principles and ethical frameworks, AI development can be aligned with societal values, promoting fairness, transparency, and accountability in all aspects of system design and implementation. To analyze the alignment of ethical theories with the proposed computing principles, we present summary tables (Table 2 and 3).

| Ethical Theory | H1 | H2 | H3 | H4 | H5 | H6 | H7 | H8 | H9 | H10 | K1 | K2 | K3 | S1 |
|---|---|---|---|---|---|---|---|---|---|---|---|---|---|---|
| Utilitarianism | C | C | C | C | C | D | C | C | C | C | C | C | C | C |
| Deontological Ethics | D | C | C | C | C | D | D | D | D | D | D | D | D | D |
| Virtue Ethics | D | C | C | D | D | D | D | D | D | D | D | D | D | D |
| Ethics of Care | D | C | C | D | D | D | D | D | D | D | D | D | D | D |
| Rights-Based Ethics | D | N | N | C | C | D | D | D | D | D | D | D | D | D |
| Social Contract Theory | D | C | C | N | C | C | C | C | C | C | C | C | C | C |
| Rawlsian Justice | D | C | C | N | C | C | C | C | C | C | C | C | C | C |
| Natural Law Theory | D | C | C | C | D | D | D | D | D | D | D | D | D | D |
| Environmental Ethics | I | I | I | D | D | D | D | I | N | N | N | N | N | N |
| Pragmatism | C | C | C | C | C | D | C | C | C | C | C | C | C | C |

*Table 2 Summary of Ethical Theories and Their Support for Principles Part 1*

| Ethical Theory | S2 | S3 | S4 | S5 | S6 | S7 | S8 | S9 | S10 | S11 | S12 | S13 | S14 |
|---|---|---|---|---|---|---|---|---|---|---|---|---|---|
| Utilitarianism | C | C | C | C | C | C | C | C | C | C | C | C | C |
| Deontological Ethics | D | D | D | D | D | D | D | D | D | D | D | D | D |
| Virtue Ethics | D | D | D | D | D | D | D | D | D | D | D | D | D |
| Ethics of Care | D | D | D | D | D | D | D | D | D | D | D | D | D |
| Rights-Based Ethics | D | D | D | D | D | D | D | D | D | D | D | D | D |
| Social Contract Theory | C | C | C | C | C | C | C | C | C | C | C | C | C |
| Rawlsian Justice | C | C | C | C | C | C | C | C | C | C | C | D | C |
| Natural Law Theory | D | D | D | D | D | D | D | D | D | D | D | D | D |
| Environmental Ethics | N | N | N | N | N | N | N | N | N | N | N | N | N |
| Pragmatism | C | C | C | C | C | C | C | C | C | C | C | C | C |

*Table 3 Summary of Ethical Theories and Their Support for Principles Part 2*

The table uses abbreviations to indicate the level of support each theory provides for the principles: D denotes direct support, where the theory fully aligns with the principle; C represents conditional support, indicating that the theory supports the principle under specific conditions; I signifies indirect support, where the theory supports the principle through broader implications; and N stands for neutral, meaning the theory neither supports nor opposes the principle.

For example, Utilitarianism conditionally supports principles such as avoiding harm to individuals (H1) and protecting organizational interests (H2) when these actions maximize overall societal well-being. Similarly, Deontological Ethics directly supports principles like transparency (H7) and respect for autonomy (H9), as these align with moral duties of honesty and individual rights. For a detailed breakdown of the reasoning behind these alignments, readers are encouraged to refer to [18].

## 5. An Approach towards Ethical Scoring

In this section, we introduce a structured methodology for evaluating the ethical robustness of data harnessing processes in the development of LLMs The ultimate objective is to establish an Ethical Risk Scoring (ERS) mechanism capable of quantifying normative risks embedded within the data pipeline. The first step involves identifying a core set of ethical dimensions that are relevant to specific data practices. Each dimension is evaluated using a set of state variables indicative of ethical performance within that dimension. These variables are integrated using a monotonic scoring function. The function must be designed to mechanistically represent the underlying logic of the ethical framework. It will also include corresponding weight sets as parameters, which encode the normative priorities of the selected ethical lens. Together, the function and the weight architecture provide a transparent and adaptable framework for translating ethical principles into quantitative assessments. Below, we present a sample scoring function and an illustrative ad hoc weight set. The resulting score should be scaled to reflect the overall ethical quality or risk across all dimensions.

### 5.1. Ethical Risk Scoring (ERS)

The questions in our framework are guided by the ethical principles outlined in Table 1. The overall structure is inspired by ten foundational ethical theories. This alignment ensures that the ethical dimensions we evaluate—such as harm, transparency, rights, and sourcing—are conceptually grounded in established ethical theories.

The ERS dimensions are: i) ethical sourcing of data, ii) transparency, iii) harm to all, and iv) respect for the rights of the data target. The state variables are designed to reflect the core intent of the dimension. We have designed a question set to test each. For this, we label the state variables as $Q_{x,y}$—e.g., $Q_{1.2}$. Here, $x$ in subscript represents the four ethical dimensions— 1 for ethical sourcing, 2 for transparency, 3 for harm potential, and 4 for target rights—while *y* denotes the state variable's order within that dimension.

To calculate the ethical risk score, the auditor must answer a set of carefully designed questions. These questions assess the state variables that influence the ethical risks, such as whether the data pertains to humans, animals, infrastructure, or the environment; whether permissions for data use were appropriately acquired; and whether safeguards against harm have been implemented. The responses to these questions are used to determine state variable values, which serve as inputs for calculating individual scores for each category: harm potential, ethical sourcing, transparency, and target rights compliance.

Using the ethical and technical concerns identified in recent literature on AI and machine learning, we developed a set of guiding questions for our ethical evaluation framework. Table 4 lists the questions, their 'yes' and 'no' answers, the corresponding state variable values, and the associated minimum and maximum weights. Questions $Q_{1.1}$ to $Q_{1.6}$ [7,9,13,16,17] represent ethical sourcing, which evaluates the ethical acquisition of data by assessing compliance with ownership rights, proper authorization, and permissions for usage. The presence of attribution mechanisms and adherence to licensing terms is also factored into the score. Questions $Q_{2.1}$ to $Q_{2.4}$ [8,10,14,17] represent transparency, which is evaluated based on safeguards against harm, validation of advertised benefits, and documentation of biases and limitations. Each criterion is assigned a weight proportional to its impact on ethical accountability. Questions $Q_{3.1}$ to $Q_{3.8}$ [5,6,8,9,15] represent harm potential of the data, which is assessed by determining whether it pertains to humans, personally identifiable information (PII), or non-human entities.

Here, we added all the ethical sourcing questions except $Q_{1.1}$ and multiplied the summation with Q1.1, as without confirming data acquisition, all other questions in this metric become invalid. For the same reason, we multiplied $Q_{4.1}$ with the summation of $Q_{3.2}$, $Q_{3.3}$, and $Q_{3.4}$, $Q_{3.5}$, and $Q_{3.6}$—because without confirming the privacy of the target, these harm-related questions become invalid for calculating the harm potential metric. Similarly, the second part of the harm potential equation depends on whether the training data includes information about non-human entities. To calculate the target rights metric, we multiply the sum of $Q_{3.2}$ and $Q_{3.3}$ with the sum of the target rights questions, since without human-related or sensitive data, the target rights questions become invalid.

The scoring accounts for intentional and unintentional harm, as well as harm to other entities such as animals, infrastructure, and the environment. A weighted monotonic function is used to calculate the harm score based on responses to questions about potential misuse or inadvertent harm. Questions $Q_{4.1}$ to $Q_{4.5}$ [9,13] represent target rights, which ensure respect for data subjects' rights, including informed consent, privacy, and the ability to opt out. Higher weights are assigned to critical rights such as data usage consent and revocation mechanisms. The final ethical risk score is computed by summing these four individual scores. The formalized logical equations for each ethical metric, along with the final score equation, are as follows:

Ethical Sourcing (S) = $\{Q_{1.1} \times (Q_{1.2} + Q_{1.3} + Q_{1.4} + Q_{1.5} + Q_{1.6})\}$  (1)

Harm Potential (H) = $\{(Q_{3.2} + Q_{3.3}) \times Q_{4.1} \times (Q_{3.4} + Q_{3.5} + Q_{3.6})\} + \{Q_{3.1} \times Q_{4.2} \times (Q_{3.7} + Q_{3.8})\}$  (2)

Transparency (T) = $(Q_{2.1} + Q_{2.2} + Q_{2.3} + Q_{2.4})$  (3)

Target Rights (R) = $\{(Q_{3.2} + Q_{3.3}) \times (Q_{4.1} + Q_{4.2} + Q_{4.3} + Q_{4.4} + Q_{4.5})\}$  (4)

Ethical Risk Score (ERS) = $S + H + T + R$  (5)

This cumulative score provides a comprehensive measure of the ethical risks associated with the data harnessing process. By identifying and addressing weaknesses in any of the four scoring categories, stakeholders can ensure the process aligns with ethical principles and proactively mitigates potential harms. This structured approach integrates ethical accountability into the core of LLM development, facilitating the creation of systems that prioritize both performance and responsibility.

| Tag | Question | Ans. | Max Value | Ans. | Min Value | Alpha Ltd. | Beta Ltd. |
|-----|----------|------|-----------|------|-----------|------------|-----------|
| $Q_{1.1}$ | Has the data been acquired? | Yes | 1 | No | 0 | Yes | Yes |
| $Q_{1.2}$ | Has the data been acquired properly from the owner? | No | 0.5 | Yes | 0 | Yes | No |
| $Q_{1.3}$ | Has it been verified that the owner has proper authorization to distribute the data? | No | 0.25 | Yes | 0 | Yes | No |
| $Q_{1.4}$ | Has the owner provided usage permission for the purpose of model training? | No | 0.25 | Yes | 0 | Yes | No |
| $Q_{1.5}$ | Has the owner provided usage permission for the application of the trained model? | No | 0.25 | Yes | 0 | Yes | No |
| $Q_{1.6}$ | If the owner required attribution, does an attribution mechanism exist and is it being used? | No | 0.25 | Yes | 0 | No | No |
| $Q_{2.1}$ | Are there sufficient process safeguards to ensure no harm to humans? | No | 2 | Yes | 0 | Yes | No |
| $Q_{2.2}$ | Are there sufficient process safeguards to ensure no harm to other entities (animals, environment, infrastructure, etc.)? | No | 1 | Yes | 0 | Yes | No |
| $Q_{2.3}$ | Have the claimed/advertised benefits of the data been sufficiently validated and documented? | No | 0.1 | Yes | 0 | No | Yes |
| $Q_{2.4}$ | Have bias and other harmful impacts of the data been documented and disclosed? | No | 0.15 | Yes | 0 | No | Yes |
| $Q_{3.1}$ | Is there data other than related to human or individual (non-human)? | Yes | 0.5 | No | 0 | Yes | Yes |
| $Q_{3.2}$ | Is there data related to human (Human)? | Yes | 1 | No | 0 | Yes | Yes |
| $Q_{3.3}$ | Is there human related data that also identifies individuals (PII)? | Yes | 0.5 | No | 0 | Yes | Yes |
| $Q_{3.4}$ | Will this data be used to harm the subject? (principle of no harm - intention)? | Yes | 0.2 | No | 0 | Yes | Yes |
| $Q_{3.5}$ | Can this data be used to harm other human beings? (principle of no harm - intention)? | Yes | 0.8 | No | 0 | No | Yes |
| $Q_{3.6}$ | Can this data be inadvertently used to harm the subject or HUMAN (unintentional)? | Yes | 0.25 | No | 0 | No | No |
| $Q_{3.7}$ | Can this data be used to critically harm other entities (animals, environment, infrastructure, etc.) through ML? | Yes | 0.5 | No | 0 | No | No |
| $Q_{3.8}$ | Can this data be used to harm other entities (animals, environment, infrastructure, etc.) through ML? | Yes | 0.25 | No | 0 | No | No |
| $Q_{4.1}$ | Does the process conform to the privacy asked by the target? | No | 0.15 | Yes | 0 | Yes | Yes |
| $Q_{4.2}$ | Was the target informed, and did they consent to the method of data collection? | No | 0.2 | Yes | 0 | No | No |
| $Q_{4.3}$ | Was the target informed, and did they consent to the process of data archiving? | No | 0.1 | Yes | 0 | No | Yes |
| $Q_{4.4}$ | Was the target informed, and did they consent to the data distribution? | No | 0.15 | Yes | 0 | No | No |
| $Q_{4.5}$ | Does the target have a means to change mind and opt out? | No | 0.15 | Yes | 0 | No | No |

*Table 4 Question set for Ethical Harnessing Metrics*

## 5.2.    Design Choices

The ethical dimensions, evaluative questions, and corresponding initial weightings presented herein are derived from the normative principles and theoretical constructs established in this study. The current configuration has been deliberately engineered as a demonstrative prototype, offering a modular and extensible foundation for ethical scoring systems. Domain experts can extend these sets with additional context-specific ethical priorities for real-world applications, framing ethical tests through the perspectives of these theories, with weights determined by theoretical consensus to enhance acceptability in a world that honors diverse ethical norms.

## 5.3.    Scoring Example

To demonstrate the application of the ERS index, we analyze two hypothetical LLM developers.

A) Alpha Ltd., a healthcare company, has developed a large language model (LLM) to assist with patient care, diagnostics, and research. The LLM is trained on diverse datasets, including web-scraped content, patient medical records, academic publications, healthcare reports, and clinical trial data. Permissions were obtained specifically for treatment and research purposes, but not explicitly for model training. Data preprocessing includes tokenization, removal of incomplete records, biased content, and repetitive PII data. Deduplication is minimal, and although the training data is updated periodically, there is no robust policy for data archiving or distribution.

B) Beta Ltd. has developed an LLM designed to provide academic and career advice to university students by analyzing vast datasets, including publicly available academic journals, career guides, university reports, and online forums. While some proprietary data from university career centers is included, the company does not implement advanced privacy measures or bias mitigation techniques, such as obtaining proper authorization and consent from data owners and subjects for model training. The LLM's recommendations are influenced by its reliance on general, unverified data sources, raising concerns about accuracy, fairness, and reliability. Additionally, the lack of rigorous safeguards and regular vulnerability assessments makes the system susceptible to attacks, potentially compromising the ethical harnessing of its data and recommendations.

The rightmost two columns of Table 4 present the hypothetical answers to the ERS questions for Alpha Ltd. and Beta Ltd. Alpha Ltd. received a lower ethical risk score of 1.4, while Beta Ltd. received a higher score of 8.4 (shown in table 5), indicating greater ethical concerns in its data harnessing process.

| Metrics | Alpha Ltd. | Beta Ltd. |
|---|---|---|
| Ethical Sourcing | 0.25 | 1.5 |
| Harm Potential | 0.25 | 3 |
| Transparency | 0 | 3 |
| Target Rights | 0.9 | 0.9 |
| Ethical Risk Score | 1.4 | 8.4 |

*Table 5 Metrics values of Alpha Ltd. and Beta*

## 6. Conclusions

The ethical challenges associated with the development of LLMs highlight the pressing need for robust frameworks that prioritize responsible data harnessing. To support actionable and quantifiable assessment of ethical risks, we propose the Ethical Risk Scoring (ERS) framework. This evaluation system is based on a structured set of assessment questions, each supported by a set of underlying ethical premises or principles. These principles are further strongly supported by major ethical theories. Evaluating a domain-specific principle through the lens of a given ethical theory is inherently non-trivial due to the depth, nuance, and internal complexities of these frameworks. Future research is needed on evaluation methodologies to systematically compare and interpret ethical theories in relation to specific principles and applied contexts. An important question arises: how are answers to these ethical evaluation questions obtained? In the current framework, we assume the involvement of a human auditor—an approach aligned with prevailing practices in many real-world scoring systems [19]. However, in the context of AI agents, new ethics technologies may increasingly play a central role. Such innovations have the potential to not only advance the robustness of ethical evaluations but also enable holistic, autonomous self-compliance mechanisms embedded within AI systems. As LLMs continue to evolve, integrating ethical safeguards will be essential to balancing technological advancement with societal well-being, paving the way for AI systems that are not only powerful but also principled. Such muti-theoretic ethics framework empowers developers, policymakers, and other stakeholders to proactively address ethical vulnerabilities, fostering trust, transparency, and global scale sustainability in LLM applications.